# Geometry helps in routing scalability

**Matt Piekenbrock**
piekenbrock.m@northeastern.edu
Khoury College of Computer Sciences, Northeastern University

## Abstract

Delay Tolerant Networking (DTN) aims to address a myriad of significant networking challenges that appear in time-varying settings, such as mobile and satellite networks, wherein changes in network topology are frequent and often subject to environmental constraints. Within this paradigm, routing problems are often solved by extending classical graph-theoretic path finding algorithms, such as Dijkstra's or $A^*$, to the time-varying setting; such extensions are simple to understand, but they have strict optimality criteria and can exhibit non-polynomial scaling. Acknowledging this, we study time-varying shortest path problems on *metric graphs* whose vertices are traced by semi-algebraic curves. As an exemplary application, we establish a polynomial upper bound on the number of topological critical events encountered by a set of $n$ satellites moving along elliptic curves in low Earth orbit (per orbital period). Experimental evaluations on networks derived from STARLINK satellite TLE's demonstrate that not only does this geometric framework allow for routing schemes between satellites requiring recomputation an order of magnitude less than graph-based methods, but it also demonstrates *metric spanner* properties exist in metric graphs derived from real-world data, opening the door for broader applications of geometric DTN routing.

## 1 Introduction & Motivation

Routing data efficiently between nodes in a network is fundamental to communication. A recurring challenge encountered by many communications networks is the ability to maintain a high transmission throughput and low resource utilization in the presence of adverse conditions brought on by the physical environment [1]. For example, certain environments have extremely low signal-to-noise ratio (SNR), high doppler shift, and long latency, requiring transmission nodes to use spectrum efficient communications or draw atypical amounts of power. The space environment poses many such unique challenges, including e.g. a lack of end-to-end connectivity; possible disruption due to the enclosing radiation environment; lack of a fiber-optic "backbone" network; an ever-changing network topology; and latency on the order of minutes. Indeed, the propagation delay of signals in deep-space alone may preclude the use of traditional Internet protocols, such as the Transfer Control Protocol (TCP), which are largely based on an effectively instantaneous flow of information between sender and receiver nodes.

One solution proposed to address these myriad of challenges is Contact Graph Routing (CGR), which is a "dynamic routing system that computes routes through a time-varying topology of scheduled communication contacts in a network" [2]. Based on the more general *Delay-tolerant networking* (DTN) architecture, CGR focuses on scaling routing problems in satellite network communication networks by exploiting the fact that transmission opportunities —or *contacts* —between satellites (or other transmitting entities in space) are generally known in advance due to the high-precision[1] location information available from orbital propagators. These contacts enable the use of traditional shortest-path (SP) algorithms on *contact graphs*, which are essentially *s-t* path-induced subgraphs of the larger satellite network representation. The benefit of the contact-graph representation is a simplifying one: by changing the graph representation of the network to account for time, one reduces the daunting task of computing SPs in a time-varying graph to the computation of SPs in a particular static representation, an easily handled task. Though elegant, one drawback to this approach is that it is inherently combinatorial: the size of the static representation depends completely on how frequently the network is changing, which in many applications can be exponential (or worse).

In pure graph-theoretic settings, computing SPs optimally can incur significant time and space costs. For example, for general *n*-vertex graphs, any shortest-path routing scheme requires a routing table size of at least $\Omega(n^{7/6})$ bits on each node [4]. The situation becomes more dire in the time-varying setting: if it exists, the Kleene star $D^*$—an encoding of all the shortest path trees—of a weighted graph $G = (V, E)$ with $n = |V|$ and $k < n(n-1)/2$ independently[2] varying edge weights takes $O(k \cdot 2^k \cdot n^3)$ time to compute [5] and requires the routing table at every node have size proportional to the number of such trees. Indeed, the existence of $O(2^k)$ pairwise incomparable shortest path trees in general precludes a polynomial bound on both the storage and computation of optimal routes per-node in the time-varying setting.

One way to overcome strict size constraints is to incorporate *metric structure*. On graph-related problems, incorporating geometry often provides opportunities to reduce the computational complexity of certain operations, or to enable similar operations to be $(1+\epsilon)$-approximated—see [6] for an overview. In the context of routing, adding metric structure to the routing scheme provides an avenue to shift the pathing strategy from a global to a local one: whereas many well-established shortest-path algorithms require (potentially) full knowledge of network topology to maintain the optimality of the corresponding paths, local routing algorithms use only knowledge of the locations of the source, target, and neighborhood nodes.

In what follows, we discuss a variety of ways to tackle the routing problem in time-varying settings. Namely, we review a few time-varying graph representations, discuss ways of transforming between them, and outline the associated complexities of SP-related problems on each graph type. Inspired by their use in large scale ad-hoc networks, we show one example of how to move the routing problem from the purely combinatorial setting to a more geometric one, illustrating the advantages that come with such a transition. Finally, we show experimental results demonstrating that the incorporation of such theory does indeed result in a more scalable routing scheme in real-world settings.

## 2 Background & Notation

Let $G = (V, E)$ be a graph with a finite vertex set $V$ and edge set $E \subseteq V \times V$. unless otherwise state, we assume $G$ is simple. In this work, $G$ may be undirected or directed depending on the context; we use $(u,v) \in E$ to denote an edge in the former and $u \to v$ to denote a directed edge in the latter. Graphs often come with additional structure in the form of real-valued weight functions $w : \mathcal{X} \to \mathbb{R}$ equipped to either the vertex set ($\mathcal{X} = V$) or the edge set ($\mathcal{X} = E$); in either case, we call $G$ a *weighted graph*. A *path* in $G$ along vertices $v = v_1, ..., v_k = v'$ is a sequence of edges satisfying $(v_i, v_{i+1}) \in E$ for all $1 \leq i < k$, which we write as $(v \rightsquigarrow v')$ for brevity sake.

---

[1] The uncertainties of orbital determination is beyond the scope of this work, see [3] for an overview.
[2] This bound assumes the variable edge weights are *separated*.

$$v \rightsquigarrow v' \quad \Leftrightarrow \quad \{(v_i, v_{i+1}) \in E : v = v_1, v_2, ..., v_k = v' \text{ for all } 1 \leq i < k\} \tag{1}$$

Unless otherwise stated, we assume a path may be *non-simple*, i.e. a path may contain repeated vertices. The *length* of a path $\ell(v \rightsquigarrow v')$ is defined as the number of edges participating in the path, and the *weight* of a path $\omega(v \rightsquigarrow v')$ is the sum of the weights of its edge weights. The *shortest-path distance* $\pi(v \rightsquigarrow v')$ between two vertices $v, v' \in V$ is the minimal weight of any path $v \rightsquigarrow v'$; we denote *shortest-path* itself with $\left(v \overset{\pi}{\rightsquigarrow} v'\right)$, i.e. $\pi(v \rightsquigarrow v') = \omega\left(v \overset{\pi}{\rightsquigarrow} v'\right)$. We say a graph is *unweighted* if the weight of every edge $v \to v'$ in $E$ is 1, in which case we have $\omega(v \rightsquigarrow v') = \ell(v \rightsquigarrow v')$ for all $(v, v') \in E$. A subgraph $G' \subseteq G$ is called a *spanner* of $G$ if $\pi_{G'}(v \rightsquigarrow v') \leq t \cdot \pi_G(v \rightsquigarrow v')$ for all pairs of vertices $v, v' \in V$, for some constant $t \geq 1$. The factor $t$ is called the *stretch factor* of $G'$ and the graph $G'$ a *t-spanner* of $G$.

## 2.1 Shortest-path problems

The *shortest-path problem* and its many variants are among the most well-studied of the classical graph theoretic problems. In this setting, we are given a graph $G = (V, E)$, a weight function $w : E \to \mathbb{R}$, and possibly some subset of vertices $V' \subseteq V$ with which we want to compute paths in $G$ of minimal length. There are three common variants:

(SP) Calculate the "*s-t*" shortest-path, if it exists, from source node $s$ to target node $t$

(SPT) Calculate the *shortest-path tree* from a source node to every other node

(APSP) Calculate the shortest-path between *all pairs of nodes*

Solutions to these variants can be used to solve many similar problems: computing the shortest-path from every node to a target node $t$ can be solved by reversing the edges and reducing to SPT; computing the *longest-path*[3] can be solved by reversing the signs on the edge weights and reducing to SP; computing the shortest-path including vertex weights $w : V \to \mathbb{R}$ reduces to SP by adding edges at every $v \in V$, and so on.

Both the effective performance and the complexity of many SP problems depends greatly on both the structure of $G$ and the distribution of edge weights—we recall the main results. Given a graph $G = (V, E)$, let $n = |V|$ and $m = |E|$. For simplicity, assume $m \sim \Omega(n)$. If $G$ is a directed acyclic graph (DAG), the SP problem is easily solved in $O(m + n)$. For general graphs $G$ (directed or undirected) with $w : E \to \mathbb{R}_+$, (SP) can be solved easily with Djikstra's algorithm [7] in $O(m \log n)$ time with a simple priority queue or $O(m + n \log n)$ using Fibonacci heaps [8]. If $G$ is directed with $w : E \to \mathbb{R}$ and contains no negative cycles, Djikstra's cannot be used; the complexity of solving (SP) jumps to $O(nm)$ using the algorithm by Bellman, Ford, and Moore [9]. If $G$ contains negative cycles, no non-simple shortest-paths exists—computing the *shortest-simple-path* reduces to the Hamiltonian path problem, which is NP-hard [10].

As the reduction to the Hamiltonian implies, the presence of negative weight cycles presents non-trivial barriers to the computation of shortest-paths. To illustrate this, assume we have a directed graph $G = (V, E)$ with no negative-weight cycles. A well-known fact in graph theory is that the lengths of shortest-paths respect the triangle inequality:

$$\ell\left(v_i \stackrel{\pi}{\leadsto} v_j\right) \leq \ell\left(v_i \stackrel{\pi}{\leadsto} v_k\right) + \ell\left(v_k \stackrel{\pi}{\leadsto} v_j\right) \quad \forall\, i, j, k \in [n] \tag{2}$$

---

[3]Note this is not to be confused with the canonical *longest-path* path is to calculate the longest *simple* path, which is NP-hard.

A more general principle holds sometimes referred to as *Bellman's Principle* [11]. In the context of SP-type problems, we give a specialized version of Bellman's principle next.

**Definition 2.1.1** (Optimal substructure): Let $G = (V, E)$ denote a directed graph equipped with edge weights $w : E \to \mathbb{R}^-$ with no negative-weight cycles, and let $\pi_k(v \leadsto v')$ denote the shortest-path distance from $v$ to $v'$ using at most $k$ steps, where $\pi_k(v \leadsto v') = \infty$ if no such path exists. For some fixed source vertex $s \in V$, a SP problem is said to have *optimal substructure* if for all $k \in [n]$:

$$\begin{aligned}\pi(s \leadsto s) &= 0 \\ \pi_{k+1}(s \leadsto v) &= \min_{u \neq v}(\pi_k(s \leadsto v), \pi_k(s \leadsto u) + w_{uv})\end{aligned} \tag{3}$$

Similarly, an APSP problem on a graph $G$ is said to have optimal substructure if every path $\pi(u \leadsto v)$ for any $u, v \in V$ has optimal substructure.

Thus, in a SP problem that has optimal substructure, subpaths of shortest-paths are shortest-paths themselves. To demonstrate this, consider the follow small example network shown below:

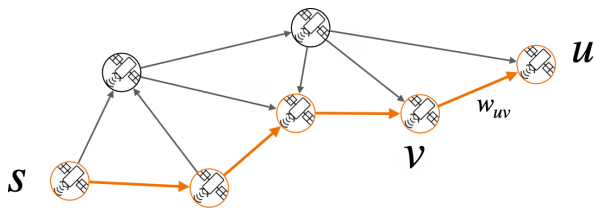

Figure 1: Example *optimal substructure* property for a single path $s \stackrel{\pi}{\leadsto} u$; the property requires that if $s \stackrel{\pi}{\leadsto} u$ is a shortest path, then $s \stackrel{\pi}{\leadsto} v$ is as well, thus $\pi(s \leadsto u) = \pi(s \leadsto v) + w_{uv}$.

The optimal substructure property asserts that the path $s \leadsto v$, which is contained in the shortest path $s \leadsto u$, is itself a shortest path. Note that the assumption of no negative cycles is a necessary assumption for an SP to have optimal substructure, and that optimal substructure is itself a *necessary condition* for optimality: if the assumption doesn't hold, then the optimality of the shortest-paths is not guaranteed. Examples of algorithms which either implicitly or explicitly assume optimal substructure to ensure optimality include Djikstra's (SPT), Floyd-Warshall (APSP), and the famous Edmonds-Karp maximum-flow algorithm [12]. To demonstrate the benefits this principle provides to path algorithms, we give a simple example showing connections between the APSP problem and matrix multiplication next.

*Example*: Consider the directed graph $G = (V, E)$ with $n = |V|$ vertices and $m = |E| \subseteq V \times V$ edges equipped with a weight function $w : E \to \mathbb{R}$ that has no negative-weight cycles. We can re-phrase the APSP problem using matrices; let $A \subset \mathbb{R}^{n \times n}$ denote the weighted adjacency matrix of $G$, where:

$$A[i, j] = \begin{cases} 0 & i = j \\ w(i, j) & (i, j) \in E \notin E \\ \infty & (i, j) \end{cases} \tag{4}$$

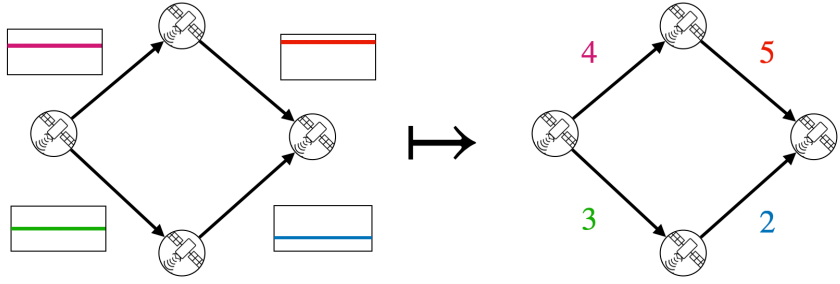

Figure 2: Illustration of how the Parameterized Shortest Path problem naturally generalizes the Shortest Path problem via the constant function $w(\lambda, e) = c_e$.

If one replaces[4] the typical addition/multiplication operations $(+, \times)$ used in traditional matrix multiplication with $(\min, +)$ operators—*min-plus matrix multiplication*—the shortest-path distances obtainable in *k*-hops can be expressed with matrix powers:

$$(A^k)[i,j] \quad \Leftrightarrow \quad \pi_k(v_i \rightsquigarrow v_j) \tag{5}$$

It can be shown ([13]) that $A^{n-1} = (a_{ij})$ is a matrix of shortest path lengths *if and only if*:

$$\begin{aligned} a_{ii} &= 0 \\ a_{ij} &\leq a_{ik} + a_{kj}, \quad \forall\, i, j, k \in [n] \end{aligned} \tag{6}$$

The connection back to the optimal substructure definition Equation 3 should be clear.

It turns out this way of looking at SP problem leaves naturally to an elegant formulation the *parameteric SP problem* (PSP) [5, 14]. In the PSP, we have as input a directed graph $G = (V, E)$ and, in addition to weight function $w : E \to \mathbb{R}$ on the edges, we have a *parameterized* weight function:

$$w(\lambda, e) = w(e) - \lambda, \quad \text{for all } \lambda \in \mathbb{R} \tag{7}$$

Some authors define the weight function over a subset of edges $E' \subseteq E$ (e.g. [15]), though this setup can be reduced to the former by setting $w(\lambda, e) = w(e)$ for all $e \in E'$ and $\lambda \in \mathbb{R}$. Note that due to Equation 7, the shortest path distances from any fixed node $s \in V$ satisfy:

$$\pi(\lambda)(s \rightsquigarrow u) \leq \pi(\lambda)(s \rightsquigarrow v) + w(\lambda, v \to u) \tag{8}$$

for every edge $v \to u$, with equality if the edge $u \to v$ is in the shortest path tree of $s$. The reader may compare Equation 8 with Equation 3. The (PSP) can be solved in $O(nm + n^2 \log n)$ time [16] on a graph with *n*-vertices and *m* edges.

Suppose the function equipped to each edge is constant for all values of $\lambda$, i.e. $w(\lambda, e) = c_e$ for some fixed constant $c_e$ that depends only on the edge. Observe the (PSP) reduces immediately to (SP), see Figure 2. Combining this observation with Equation 8, one might argue that the PSP is a kind of natural way to expand the SP* family of problem to the parameterized setting. Indeed, this elegant interpretation has more extensive connections to the field of tropical geometry. The curious reader is referred to [5] for more details.

Problems satisfying the optimal substructure property are typically solved *optimally* by employing a combination of dynamic programming or greedy-like algorithms. This motivates the following natural question: "When conditions are needed for a problem to admit a greedy algorithm that yields optimal solution(s)?" It turns out this problem can be readily answered using Matroid theory [17], or more generally through the use *independence-sets*. Greedily solving a problem using a Matroid-based approach turns out to be not only a sufficient condition for optimality, but also a necessary one: if a problem can be solved by greedy algorithm optimally, it has to be a Matroid. Since our focus is the time-varying setting and the theoretical background on this topic is deep, we do not delve into this further; the interested is reader is referred to [17–21] for more details.

---

[4]The replacement of $(+, \times)$ with $(\min, +)$ is sometimes referred to as the tropical version of matrix multiplication.

## 3 The Temporal Setting

We now consider an intuitive and attractive variant of SP* family of problems called the *Time dependent shortest-path* (TDSP) problem, which may be interpreted as a particular analog of the SP problem in the continuous time-varying setting. Let $G = (V, E, T)$ denote a time-varying graph over the a fixed *time horizon* $[0, T]$ where each edge is equipped with a weight function $w : E \times \mathbb{R} \to \mathbb{R}_+$, parameterized by *time*. For simplicity, we use $w_{ij}(t)$ to denote this function for the edge $(i, j) \in E$. A common interpretation for $w_{ij}(t)$ is the *travel time* along edge $(i, j)$.

**Definition 3.1** (TDSP): Given a graph $G = (V, E, T)$ over a time horizon $[0, T]$ and a time-dependent weight function $w : E \times \mathbb{R} \to \mathbb{R}_+$, the *time-dependent shortest path* problem is to find the shortest *s-t* path minimizing:

$$\text{TDSP}(s \rightsquigarrow t, t_0) = \min_{s \rightsquigarrow t} \sum_{i=1}^{k-1} w_{ij}(t_i), \quad t_{i+1} = t_i + w_{i,i+1}(t_i), \text{ for all } i \in [k] \qquad (9)$$

where $w_{ij}(t)$ denotes the weight of traversing the edge $v_i \to v_j$ at time $t$, and as usual the path $s \rightsquigarrow t$ means $s = v_1, v_2, ..., v_k = t$.

Note that in this definition, by convention we begin traversing $G$ from node $s$ at time 0.

Another interpretation of the FIFO property is that messages transmitted throughout the network always traverse across nodes in a FIFO order. Some authors refer to FIFO networks as networks wherein "waiting at nodes is never beneficial"[22]. Since property is essential to the solvability of the TDSP, we give multiple interpretations of it now.

**Definition 3.2** (FIFO condition): A graph $G = (V, E)$ equipped with a weight function $w : E \times \mathbb{R} \to \mathbb{R}_+$ is called *FIFO network* if:

1. It is PL and has slope greater than $-1$ ($dw/dt > 1$)[23]
2. $t < t' \implies t + w_{ij}(t) < t' + w_{ij}(t')$ for all $(i, j) \in E$
3. The *arrival time* function $a_{ij}(t) = t + w_{ij}(t)$ is non-decreasing for all $(i, j) \in E$ [22]

To help understand the kinds of networks that respect FIFO, we give an analogy below.

*Example*: Suppose a passenger could leave an airport at 6:30 am and arrive at his intended destination 6 hours there-after. If there exists a flight that leaves 7:00 am at the same airport that has a travel time of $< 6$ hours, then such a time-varying network would *not* respect the FIFO property, as it is beneficial to the passenger to spend some amount of time waiting. In contrast, one would never arrive earlier at one's destination in a road network by pulling off the highway and sitting in a parking lot for any amount of time—thus road networks could be loosely interpreted as respecting the FIFO condition.

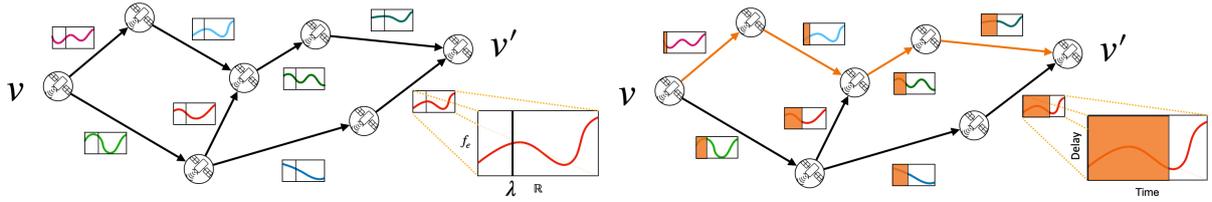

Figure 3: The parameterized SP and TDSP problems. The former computes SP variants for all values of $\lambda \in \mathbb{R}$, which reduces to the static setting when $\lambda^*$ is fixed. In the latter, the cost of a path is time-dependent as the travel-time of an edge depends on adjacency arrival-times.

We remark that, if the weight function attached to the edges is PL, then (TDSP) can actually be transformed into (PSP) via a non-trivial conversion involving *primitive and minimization breakpoints* [23]. This is an interesting observation because solving the (TDSP)Equation 9 optimally in non-FIFO networks is NP-hard [23–25], indicating the reduction is non-polynomial. The situation becomes a bit more nuanced, as there exist some cases where waiting is penalized constrained that are actually polynomial [26]. The reduction used by Foschini [23] connecting (PSP) and (TDSP) used an intermediate acyclic representation, which they referred to the *layered graph representation*. We discuss this more detail in the next section.

## 3.1 Size bounds and optimality conditions

As shown above, though the conditions under which the static SP algorithms (e.g. Djikstra's, Floyd Warshall) yield optimal results is a well studied topic, extensions of these conditions to the time-varying setting is more nuanced. Perhaps the most intuitive way to to extend the $SP^*$ family of algorithm to the time-varying setting is to try to convert a time-varying graph representation into a static representation such that one may apply any of the discussed $SP^*$ algorithms without modification. We formalize this as follows:

In what follows, let $G = (V, E, T)$ denote a discrete time-varying graph over a fixed time horizon $\mathcal{H} = [0, h]$ equipped with time points $T_e \subset \mathcal{H}$ associated to every edge $e \in E$. Moreover, let $(\mathcal{T}, <)$ denote the totally ordered set of edge times, i.e.

$$\mathcal{T} = \cup_{e \in E} T_e = \{\, t_1, t_2, ..., t_p \,\} \tag{10}$$

The "unrolled" version of $G$ with respect to $T$ is called its *static expansion*.

**Definition 3.1.1** (Static-expansion): The *static expansion* $G_\mathcal{T} = (V_\mathcal{T}, E_\mathcal{T})$ of $G$ over the index set $\mathcal{P} = [\,p\,]$ is defined as as the directed acyclic graph obtained by connecting a sequence of graphs $G_1, G_2, ..., G_p$ with vertices and edges satisfying:

1. $v \in V \implies v(i) \in V_i \subset V_\mathcal{T}$ for all $i \in \mathcal{P}$
2. $e = (\,v(i), v'(i)\,) \in E_i$ and $t_i \in T_e \implies (\,v_i, v'_{i+1}\,) \in V_i \times V_{i+1} \subset E_\mathcal{T}$

A figure depicting the static expansion of a simple time-varying graph is given below. These are sometimes called *time-extended graphs* or *time-layered graphs*. One example application of the static expansion of a time-varying graph is the reduction of a particular time-varying SP problem to a static SP problem. To illustrate one of the many useful properties of the static expansion, we show a toy problem with the following network.

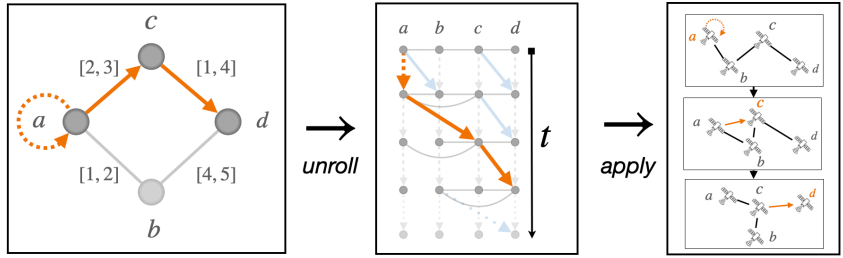

Figure 4: Example performing a static expansion or "unrolling" of a time-varying graph in interval representation to its static equivalent.

Here, the interval-representation of a time varying network is shown on the left, where edge edge is annotated with a set of intervals indicating edge connectivity. Suppose, in Figure 4, one wanted to route message from node *a* to node *d*, starting at time $t = 1$. Assume it takes unit time to route any given message along a single edge. There exists an edge connecting *a* to node *b* at times $t \in [1, 2]$, however one may confirm that by waiting at node *a* for 1 unit of time, one may route the message with a shorter arrival time via node *c*, requiring a SP algorithm to have the ability to wait at a given node (or equivalently, to consider self-loops). By replicating the network at each discrete step in time, one may avoid this problem while retaining the ability of using any well-established static SP algorithm (e.g. Djikstra's, Bellman Ford, etc.) with the static expansion.

One of the many nice properties of static expansion graphs is that they are inherently acyclic, implying paths between nodes are composed of at most $O(n)$ pieces. However, for arbitrary graphs the size of the nodes from $n$ to $O(n^2)$, implying the size of the static expansion could be as large as $O(T \cdot n^2)$ where $T$ is the number of discrete time steps (the 'time horizon') the expansion is applied over. If the vertices in the time-varying graph lie in $\mathbb{R}^d$ and move along according to *algebraic motion*—that is, along polynomial curves of constant maximal degree $s$ —then one can bound $T$ by $O(n^{d-1+\epsilon})$ for any $\epsilon > 0$ under periodicity assumptions [27]. This lies in stark contrast with the complexity results of Orda and Rom [28], who showed that the arrival time function could have as many as $n^{\Omega(\log n)}$ pieces for any given *s-t* path. Surprisingly, in a followup paper Foschini et al [23] showed a reduction of the computation of the arrival time function to a static-expansion graph by exploiting connections between the TDSP and the parametric SP problem; they also show that a minimum *delay* path can be computed in polynomial time. Interesting, a single parametric *s-t* problem over an *n*-vertex graph with *m*-edges can be solved in $O(nm + n^2 \log n)$ [16], though the number of SPs that may exist could be as large as $O(2^n)$ [5].

The scalability of the static expansion graph depends heavily on the scaling of the time horizon, $T$, and on the density of edges in the over-arching satellite network. For example, in CGR, the size of the multi-graph representation network is proportional to the size of the contact plan, which itself depends on the number of distinct times satellites "have visibility" with each other in the time-evolving satellite network. Whether two satellites have visibility with each other depends on a variety of complex calculations, such as whether the antennae on each satellite is oriented correctly, whether there does not exist any non-trivial occlusions or barriers between the satellites, etc. In practice, whether or not satellites can transmit to each other is requires accurate orbital dynamics information [2, 29], which is often calculated as part of a the Satellite Orbit Analysis Program (SOAP) [30].

## 3.2 Routing

Thus far, the size of a number of time-varying graph representations and the complexity of a variety of SP algorithms on said graphs has been discussed without much of any mention to network routing as all. We now formalize the connection between *routing* and *paths* in graphs.

A *routing scheme* for a pair of vertices $v, v' \in V$ on a graph $G = (V, E)$ is a rule set for producing a path $v \rightsquigarrow v'$. A *geometric* routing scheme $R$ is a routing scheme over a metric graph $G$. A scheme $R$ is said to be *c-competitive* if the lengths of all the shortest paths produced by $R$ are no more than $c$ times the metric distance between its endpoints:

$$\pi_R(x \rightsquigarrow x') \leq c \cdot d_X(x, x'), \quad \forall\, x \in X \tag{11}$$

For a fixed class of graphs $\mathcal{G}$, the *routing ratio* is largest constant $c' \geq c$ for which a routing scheme $R$ is $c$-competitive for all $G \in \mathcal{G}$ and all pairs $(v, v')$ in $G$. Note that any geometric $t$-spanner on a metric graph yields a $t$-competitive routing scheme if the routing scheme always produces shortest paths. That is, the routing ratio on a class of graphs $\mathcal{G}$ is an upper bound on the spanning ratio of $\mathcal{G}$, as the routing ratio proves the existence of a bounded-length path.

A *local routing scheme*[5] is a routing scheme wherein the path of a transmitted message (or packet) is determined using only local information at each vertex—without knowledge of the global graph topology. In particular, a routing scheme is said to be a *k-local routing scheme* if each $v \in V$ only uses knowledge of its neighbors $N_k(v)$ which are at most $k$-hops away to determine the next vertex $w \in V$ to forward the message too, + at most $O(1)$ amount of additional information. Formally, a *k-local routing algorithm* is a function:

$$\begin{aligned} R_k : V \times V \times V \times \mathcal{P}(V) &\to V \\ (v, s, t, \mathcal{N}_k(v)) &\mapsto w \end{aligned} \tag{12}$$

where $v \in V$ is current node, $(s, t)$ are the source and target nodes, respectively, and $(w, \mathcal{N}_k(v))$ are as above. Local routing schemes are often used in ad-hoc wireless sensor networks and motion planning due to their minimal energy footprint and low energy requirements [31].

It is natural to wonder what classes of graphs $\mathcal{G}$ admit competitive routing schemes, and how localized those schemes are (if at all). Some examples of local routing schemes include *greedy*, *compass*, *right-hand-rule*, and their randomized variants; examples of classes of graphs which support such routing schemes include *triangulations*, *planar straight-line graphs*, $\Theta$-*graphs*, *Gabriel graphs*, *$\beta$-skeletons*, *Delaunay graphs*, and so on. The study of spanning ratios and the competitiveness of certain routing schemes employed on them is an active field of research, see any of [32] and references therein for an overview.

## 3.3 Geometry

We are concerned here with the study of *geometric graphs*—graphs derived from spatial or metric information. Define the *metric graph* $G$ of a finite metric space $(X, d_X)$ as the complete graph with $n = |X|$ vertices with edge weights given by the pairwise distances $d_X$. If $G' \subseteq G$ is a $t$-spanner of this graph, then $G'$ is called a *geometric t-spanner of* $(X, d_X)$. Spanners are of interest in many applications as they act as sparse encodings of the underlying metric space. In particular, geometric spanners provide $O(t)$-approximations to many computational problems involving proximity queries, e.g. closest-pair, all nearest-neighbors, and shortest path [33].

---

[5]Related classes of routing schemes which are also 'local' in a specific sense are 'online' or 'memoryless' routing schemes.

The *Delaunay triangulation* (DT) is an important triangulation that has many properties and has an innumerable number of applications, see [34–37] and references therein for a survey of results. One such common application of the DT is routing in large-scale ad-hoc networks [37]. The DT can be defined in multiple ways—we give three equivalent definitions:

1. The DT maximizes the minimum angle $\angle abc$, among all other triangulations

2. The DT is the dual of the Voronoi diagram:
$$\text{Vor}(p) = \{x \in \mathbb{R}^d /  \| x - p \| \leq \| x - p' \|, \forall p' \in P\} \quad (13)$$

3. The DT is the triangulation consisting of triangles $\triangle pqr$ whose enclosing balls are empty

Given a graph $G$, we refer the intersection between the DT and $G$ as the Delaunay *subgraph* $\Delta(G)$. It is known that $\Delta(G)$ is a 2-spanner [36] of $G$ when $G$ is the complete graph, implying that shortest-paths on metric-derived $\Delta(G)$ are at most twice the length of the Euclidean or "straight-line" paths in $G$. Below is a figure depicting some of these properties of the Delaunay subgraph:

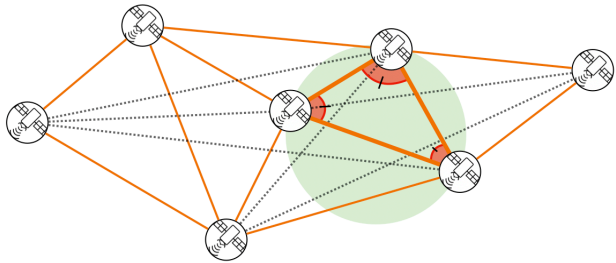

Namely, they support the use of *greedy routing* [38], which is a purely local (or "online") routing scheme. The competitiveness of local routing in the Delaunay is an active research field, see [35]. There also exist $\alpha$-stable variants of $\Delta(G)$ first presented by Agarwal et al [39], which also been studied in kinetic settings [40]—the primary setting here. In the following section, we demonstrate the usefulness of incorporating the geometric information the Delaunay subgraph provides in the context of routing.

# 4 Main Results

## 4.1 Theoretical bounds and results

We seek a polynomial bound on the number of topology-changes critical events a time-varying network may experience over a fixed time duration. To establishing this bound, we will use a tool known as a Davenport-Shinzel (DS) sequence.

**Definition 4.1.1** (Davenport Schinzel): For $n, p \in \mathbb{N}$, an $(n, p)$-Davenport Schinzel of order $p$ is a sequence $\mathcal{S} = s_1, s_2, ..., s_n$ over an alphabet $\mathcal{A}$ satisfying:

1. No two adjacent elements are equal ($s_i \neq s_{i+1}$ for all $i \in [k-1]$)

2. There is no alternating subsequence of the form ...a...b...a...b... of length $p + 2$ between any distinct symbols $a, b \in \mathcal{S}$.

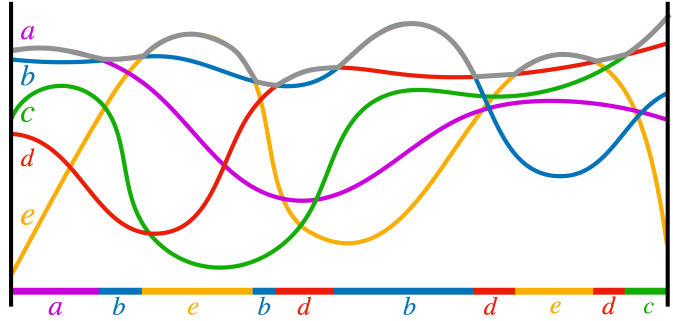

Figure 6: The upper envelope $U(\mathcal{F})$ over a set of 5 splines $\mathcal{F}$, shown above in gray. The corresponding sequence $(a, b, e, b, d, b, d, e, d, c)$ tracking the envelope is a $(5, 3)$ DS sequence.

There is a close relationship between Davenport-Schinzel sequences and the combinatorial structure of lower/upper envelopes of collections of planar curves. Namely, for any collection $\mathcal{F} = \{f_1, ..., f_n\}$ of $n$ continuous univariate functions—each pair of which intersect in at most $p$ points—the size of the upper-envelope (or lower-envelop) $U(\mathcal{F})$ is upper bounded by $\lambda_p(n)$. The upper-envelope is demonstrated in the figure below, which is composed of 5 splines, any pair of which intersect at most 4 times in the given interval.

One of the remarkable things about DS sequences is that, surprisingly, they are effectively linear [41]. Since the orbiting satellites in space follow along essentially *elliptic* orbits, orbital trajectories are a particularly convenient family of functions to study with such sequences.

Bounding the number of topological changes or "edge-flips" in the Delaunay triangulation is major area of research in computational geometry, see [40, 42] for a survey. Guibas [43] established a bound of $O(n^d \lambda_p(n))$ in 1991—essentially cubic in $n$. More recently, Rubin established a $O(n^{2+\epsilon})$ for any $\epsilon > 0$ under unit-speed motion of the vertices, and under the assumptions that (1) any four points are co-circular at most three times and (2) no triple of points can be collinear more than twice. Due to the very nice structure the time-varying satellite network exhibits relative to random positions / non-smooth trajectories associated with other dynamic environments, we also make these assumptions and adopt this bound below.

We are now ready to state our main result.

**Theorem 4.1.1**: Assume a set of $n$ satellites $S \subset \mathbb{R}^3$ are orbiting the earth are moving at unit-speed along ellipses of maximal degree 2. Moreover, assume we have $m$ fixed ground stations, and that $\Delta(S) \subseteq \text{Vis}(S)$, where $\text{Vis}(S)$ is the visibility-graph of $S$ relative to the earth. Then there exists at most $O(mn\lambda_2(n)) + (n - m)^2 \lambda_2(m)) \approx O(m \cdot n^2)$ topological critical events that change $\Delta(S)$ within a single orbital period.

*Proof.* The polynomial-time bound on $\Delta(S)$ essentially follows directly from Albers bounds on Voronoi diagram changes over time [44, 45], where the $m$ ground stations can be considered as fixed vertices in the graph. The $\approx$ statement follows from the fact that the size of that Davenport-Shinzel sequences is essentially linear for all values of degrees $s \leq 5$, scaling beyond the linear regime only by a factor that depends on $\alpha(n)$, where $\alpha(n)$ is the extremely slow-growing inverse-Ackermann function [41]. In fact, when $s = 2$, $\lambda_2(n) = 2n - 1$[46], and thus given bound follows. ∎

We would like to demonstrate the utility of this result with simulation results. Prior to doing so, we briefly describe a stabilized variant of the Delaunay triangulation introduced in

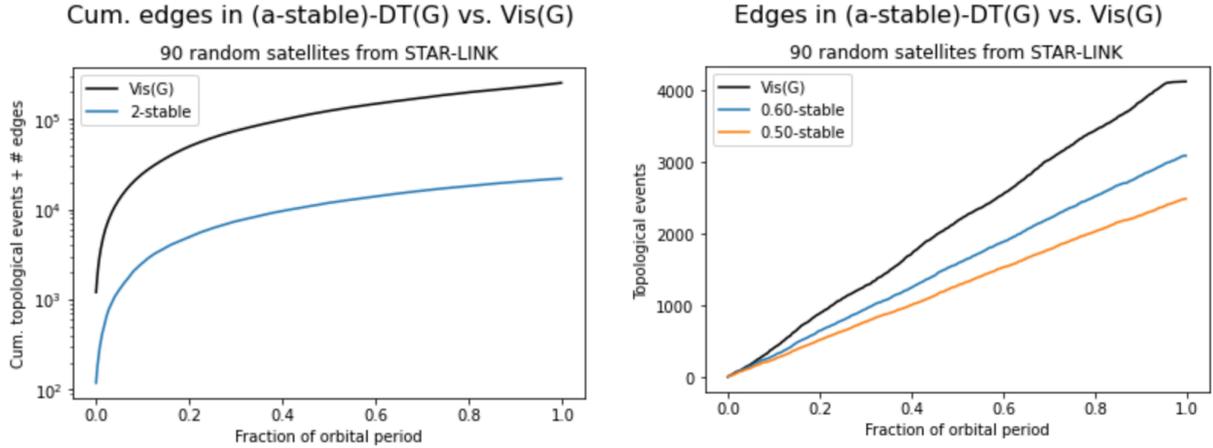

[39], which we use in the following experiments. First, observe that two triangles ($\triangle pqr^+, \triangle pqr^-$) may only exist in the Delaunay triangulation if they satisfy the following:

$$(\triangle pqr^+, \triangle pqr^-) \in \text{Del}(G) \quad \Leftrightarrow \quad \angle pr^+q + \angle pr^-q < \pi \tag{14}$$

This observation leads to a natural "stable" characterization of $\Delta(G)$, called the $\alpha$-stable Delaunay triangulation. It is given as the subgraph of $\Delta(G)$ whose edges $pq$ satisfy:

$$\angle pr^+q + \angle pr^-q \leq \pi - \alpha \ \text{ for some } \ 0 < \alpha < \pi \tag{15}$$

Thus, by varying $\alpha > 0$, we may study subgraphs of $\Delta(G)$ of varying stability and size.

### 4.2 Experiments and Empirical results

To empirically validate our findings[6], we downloaded a set two-line elements (TLEs) [47] from the Celestrak database [48] characterizing the orbital dynamics of SpaceX's "STARLINK" satellites on June 8th, 2022.

To obtain orbital positions over time from the TLEs, we use the Skyfield package [49]. Specifically, we sampled 90 random satellites and projected their orbits using these TLE's over an approximate orbital period (about 110 minutes). To get continuous representations of the dynamics, we built high-precision cubic splines tracking the distance $d_\mathcal{L}$ between an imaginary line segment $\mathcal{L}$ connecting each pair of satellites and the origin (representing the center of earth). This leads to a simplified graph representation where edges depict when pairs of satellites have line-of-sight with each other, which we denote as $\text{Vis}(t)$—the *visibility graph* at time $t \in [0, T]$. The presence of a edge in this simplified model is easily deduced by tracking the upper-envelope of $d_\mathcal{L}$ intersected with the constant function $f(t) = 6371$ (i.e. the approximate radius of the earth in kilometers). Finally, we count the number of topological events encountered by the Delaunay subgraph by a similar strategy using the closed-form *determinant* expressions for the circumcircle and orientations of a given set of points [50].

The number of topological critical events over a single orbital period—events which change the graph structure—are shown in the right in Figure 7 for both $\text{Vis}(G)$ and the $\alpha$-stable Delaunay subgraphs, for $\alpha \in \{\frac{1}{2}, \frac{4}{5}\}$. We also track, in log-scale, the cumulative number of edges within each of these graph types, which is shown on the left side of Figure 7.

---

[6]All code to reproduce the experiments performed can be found at https://github.com/peekxc/geom_dtn

As one might expect, there is an order of magnitude difference in the cumulative number of edges and topological critical events encountered by the (1/2)-stable variant $\Delta(G)$ relative to Vis($G$) over the entire time horizon. Perhaps Surprisingly, the number of critical events alone in $\Delta(G)$ appears to scale proportionally as the number of changes in Vis($G$), where the proportionality constant is determined by the choice of $\alpha$. This result is actually quite surprising, considering the conditions which determine the presence of an edge in $\Delta(G)$ and Vis($G$) are radically different—it suggests that there may always exist a stable "backbone" network whose topologically behavior is similar to that of the visibility graph. Given the strong linearity of these changes, it seems likely that this behavior will continue proportionally as the number of satellites increases, though this phenomenon needs further investigation and thus we leave it as future work.